\documentclass[aps,prb,twocolumn,superscriptaddress]{revtex4-2}
\usepackage{bm}
\usepackage{ae}
\usepackage[T1]{fontenc}
\usepackage[ansinew]{inputenc}
\usepackage{amsmath}
\usepackage{amssymb}
\usepackage{graphicx}
\usepackage{color}
\usepackage[colorlinks]{hyperref}
\usepackage{epstopdf}
\input{epsf}

\begin{document}

\title{Nanomechanical ancilla qubits generator for error correction algorithms in quantum computation}

\author{D. Radi\'{c}*}
\affiliation{Department of Physics, Faculty of Science, University of Zagreb, Bijeni\v{c}ka 32, Zagreb 10000, Croatia}

\author{L.Y. Gorelik}
\affiliation{Department of Physics, Chalmers University of Technology, SE-412 96 G{\"o}teborg, Sweden}

\author{S.I. Kulinich}
\affiliation{B. Verkin Institute for Low Temperature Physics and Engineering of the National Academy of Sciences of Ukraine, 47 Prospekt Nauky, Kharkiv 61103, Ukraine}

\author{R.I. Shekhter}
\affiliation{Department of Physics, University of Gothenburg, SE-412 96 G\"oteborg, Sweden}


\begin{abstract}
We suggest a nanoelectromechanical setup that generates properly entangled ancillary ("ancilla") qubits for error correction algorithms in quantum computing, demonstrated as an encoder for the three-qubit bit flip code. The setup is based on mesoscopic terminal utilizing the AC Josephson effect between voltage biased superconducting electrodes and mechanically vibrating mesoscopic superconducting grain in the regime of the Cooper pair box, controlled by the gate voltage. Required functionality is achieved by specifically tailored time-protocol of operating two external parameters: bias voltage and gate voltage. The superconducting grain is fixed on the free end of a cantilever, performing controlled in-plane mechanical vibrations, generating the nanomechanical coherent states organised in a pair of entangled cat-states in two perpendicular spatial directions. Cooper pair box and nanomechanical coherent states become three entangled qubits in a particular way: quantum information, initially encoded in superposition of the Cooper pair box states, is transduced into quantum superposition of two special 3-qubit entangled states, $\vert \uparrow + \, + \rangle$ and $\vert \downarrow - \, - \rangle$. It constitutes the basic input state for the three-qubit bit flip code, used in quantum computation mainly for error correction, "installed" on a single physical object in which the last two ancilla qubits are generated by the nanoelectromechanical setup. 
\end{abstract}

\maketitle

\section{Introduction}

Fault tolerant quantum computing represents one of the most challenging tasks in the field of modern quantum information (QI) \cite{Nielsen} research and technology. To achieve this goal, providing quantum processors and channels to transfer the QI between them is the basic step. However, due to interaction with environment, or due to imperfections in the fabrication process, any QI processing unit, either a processor for quantum calculations or channel for its transmission, is subjected to errors appearing in the process, which are finally reflected in inaccuracy of resulting QI. Unlike in classical computer, where this problem is solved rather easily, either by using algorithm such as "checksum" for error heralding to trigger repetition of the process, or creating multiple copies of "1" or "0" bit states and then using the "majority rule" for error correction, in quantum processing systems it is way more challenging. The main obstacle is the "no-cloning theorem" \cite{Park}, which fundamentally prohibits the very possibility of copying the state of a qubit. Qubit \cite{Schumacher} itself is a quantum system with two states where the QI is stored in their superposition \cite{Nielsen}, thus being a basic "container" of the QI. Having the "copy/paste" procedure forbidden for qubits, the other tools need to be employed. One of the most powerful among them being at hand is quantum entanglement \cite{Horodecki}. Using the property of entanglement between parts of a quantum system, different schemes, protocols or codes have been developed to address the problem of error heralding and error correction. In essence, these schemes reduce to constructing and operating the physical system in a way to entangle the operating qubit, containing the QI, to different number of ancillary, so-called ancilla qubits serving to achieve a desired functionality of error heralding or correction. Among them, a few well known examples are the 3-qubit bit flip code \cite{Peres}, for improved fidelity, or the 9-qubit Shor code \cite{Shor}, for completely random error correction, etc.
Operations upon qubits and QI stored in them are performed by logic gates, the real physical setups that correlate parts of the physical quantum system containing the QI in the specific way resulting in its desired state.
Specifically, the three-qubit bit flip code is creation of particular entangled state of three qubits, one operating and two ancillas, and then performing correction of the erroneous flipped "bit", equivalent to using five basic CNOT logic gates in a particular way. Creating the specifically entangled 3-qubit state, equivalent to using two CNOT gates, is performed by the "encoder for the bit flip code" - its nanomechanical implementation is the main topic of this paper.
With quantum information encoded in superposition of states of qubit 1 at input, i.e. $c_+ \mid \uparrow \rangle_1 + c_- \mid\downarrow \rangle_1$, the encoder provides as an output the 3-qubit state $c_+ \mid\uparrow \rangle_1 \otimes \vert + \rangle_2 \otimes \vert + \rangle_3 + c_- \mid \downarrow \rangle_1 \otimes \vert - \rangle_2 \otimes \vert - \rangle_3$ in which qubits $\mid \,\, \rangle_1$, $\mid \,\, \rangle_2$ and $\mid \,\, \rangle_3$ are entangled, where $\uparrow \downarrow$ and $\pm$ present pairs of possible qubits' states and $c_\pm$ are complex amplitudes characterising the (unitary) quantum superposition with $\vert c_- \vert^2 + \vert c_+ \vert^2 = 1$.\\

The very implementation of qubit has been a subject of extensive research, covering number of fields in physics such as optics, atomic physics or solid-state physics, as well as their combinations \cite{QuantumProcessing,Girvin,Devoret,Mirhosseini,Leibfried}, in an attempt to find an optimum between advantages and disadvantages of each implementation.
We are focused to the nanoelectromechanical (NEM) implementation in which we utilize coherent interplay of the charge-qubit \cite{Bouchiat,Nakamura,Robert,Lehnert} states and nanomechanical excitations \cite{Schneider,Hann,Chu}. The main purpose of this interplay is to achieve transduction of the QI between charge-qubit and nanomechanical subsystems. Choice of the NEM implementation brings forward a unique level of compactness, with easily electrically controlled charge-qubit (although with rather short decoherence time for the very same reason) and nanomechanics with amazingly high quality factors achievable nowadays \cite{Laird,Tao,Bereyhi}.
The NEM implementation comes down to a mesoscopic terminal based on the AC Josephson effect. It comprises a pair of voltage-biased superconducting (SC) electrodes, with a superconducting mesoscopic grain, attached to a  nanomechanical oscillator, placed amid. States in the grain are controlled by the gate voltage in the way to be in regime of the Cooper pair box (CPB), being a charge-qubit in the NEM setup. 
In our previous publications on this topic \cite{npjDR,Bahrova} we showed that dynamics of Josephson tunneling between the CPB and the SC leads resulted in formation of nanomechanical coherent states entangled with states of a charge-qubit if Josephson and mechanical frequencies are in resonance. Furthermore, applying the bias voltage manipulation protocol, we demonstrated an onset of nanomechanical cat-states consisting of coherent states.
In the later publication we showed that the QI, encoded into charge-qubit states, can be transduced into the pure nanomechanical cat-state and back, providing the time-protocol of manipulating the bias and gate voltages for that functionality \cite{PhysB_DR}. Finally, the quantum network, consisting of terminals with charge-qubits attached to the same nanomechanical vibrating beam, was suggested, demonstrating the functionality of transmission of the QI between charge-qubits, utilising entanglement to nanomechanical coherent states driven by specially tailored operating time-protocols on the SC leads and gates \cite{PhysB_DR2}.

In this paper we suggest the NEM implementation of the 3-qubit bit flip code in the terminal comprising electrically controlled pair of SC electrodes and gate electrodes with nanomechanical oscillator in the form of cantilever, with superconducting grain on top of it, performing in-plane oscillations in two orthogonal directions (see Fig. \ref{Fig-Schematics}(a)). We will show that, applying the specific time-protocol of operating the electrodes,  the QI encoded in the charge-qubit states is transduced into a pair of entangled nanomechanical cat-states, providing exactly the above-described functionality of an encoder for the 3-qubit bit flip code. In Section 2, following this introduction, we describe the physical model and present its Hamiltonian. Section 3 contains derivation of time evolution operators for specific time-protocol of operating the external control parameters to achieve desired functionality of the system.
In Section 4 we discuss some nanomechanical properties of the obtained quantum state from the physical point of view.
Final section contains discussion and concluding remarks.

\section{The Model}

\begin{figure*}
\centerline{\includegraphics[width=1.75\columnwidth]{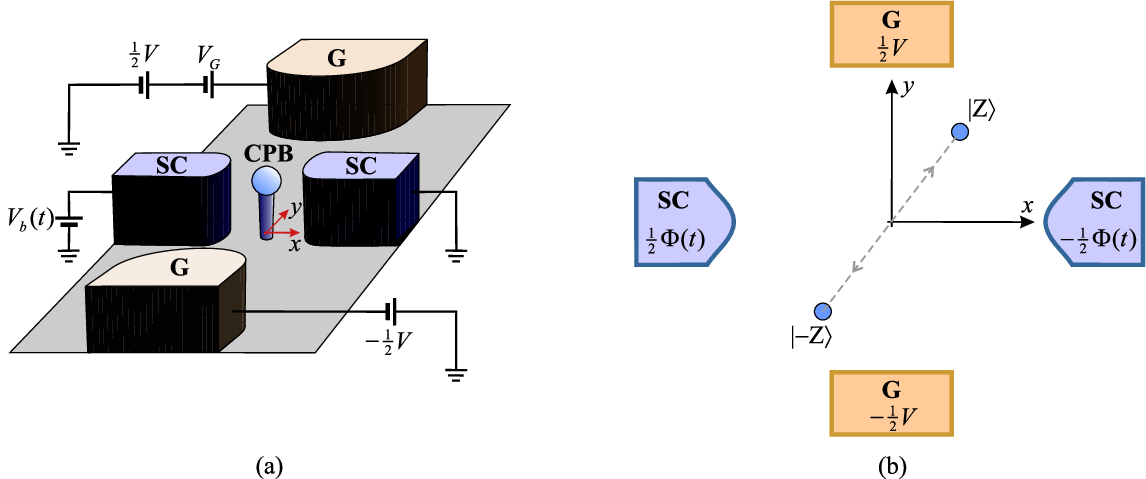}}
\caption{(a) Schematic illustration of the NEM setup featuring the AC Josephson effect. Two superconducting electrodes (SC) are voltage-biased by $V_b(t)$, controllable over time $t$, providing the superconducting phase difference $\Phi(t)$. Between them there is a cantilever with a superconducting mesoscopic grain (CPB) on the top, capable of performing the 2D mechanical vibrations in the $(x,y)$-plane. Position-dependent tunneling of Cooper pairs takes place between the CPB and SC electrodes. The role of gate electrodes (G) is twofold: (1) maintaining the state of the superconducting grain in the regime of the Cooper pair box (CPB) by constant voltage $V_G$; (2) providing an electric  field by $V$ to couple displacement to the states of the CPB in the desired way and/or to manipulate them. (b) Starting with the CPB states $\mathbf{e}_y^{\pm}$, two pairs of 2D nanomechanical coherent states, $\mid \pm Z_x(x,p_x;t) \rangle$ and $\mid \pm Z_y(y,p_y;t) \rangle$ for motion in $x$- and $y$-direction, are generated by  coherent dynamics of the NEM setup, altogether yielding two entangled states $\mid \pm Z \rangle \equiv \mathbf{e}_y^\pm \otimes \mid \pm Z_x(x,p_x;t) \rangle \, \otimes \mid \pm Z_y(y,p_y;t) \rangle$.}
\label{Fig-Schematics}
\end{figure*}

Schematic picture and description of the NEM setup is shown in Fig. \ref{Fig-Schematics}(a).
%
%
%
%
The superconducting part of the junction (SC electrodes and superconducting grain (CPB) oscillating between them) operates in the regime of the AC Josephson effect, i.e. the superconducting electrodes are biased by a constant symmetrically applied bias voltage $V_b(t)$, which can be switched on to value $V_b$ or off to zero, providing a superconducting phase difference $\Phi(t)$ dependent on time $t$. Here, $\Phi(t)=\mathrm{sgn}(V_b)\,\Omega t$ and $\Omega = 2|e|V_b/\hbar$ is the Josephson frequency. Some constant initial phase at $t=0$ is for simplicity taken to be zero. Cooper pairs tunnel between the SC-electrodes and the CPB, which is attached to a top of a "vertical" cantilever performing the 2D nanomechanical vibrations in $(x,y)$-plane at frequency $\omega$ (assuming to be equal in both directions), thus making the tunneling essentially position-dependent. Neglecting a geometric asymmetry of the junction, we expand the Josephson coupling in terms of a small parameter $\varepsilon \equiv a_0 / x_{\text{tun}} \ll 1$, where $a_0=\sqrt{\hbar/m\omega}$ is the amplitude of zero-mode oscillations ($m$ is mass of the oscillator) and $x_{\text{tun}}$ is the tunneling length.

Symmetrically placed gate electrodes have the two-fold function: (1) By the particular choice of constant potential $V_G$ on gate electrodes, the mesoscopic superconducting grain is set into regime of the Cooper pair box, i.e. the effective two-level system with degenerate states with zero and one excess Cooper pair, $\mathbf{e_y^{\mp}}$ respectively.
Those we call "the charge states" or "the CPB qubit states".
We assume that $V_G$ is always present, preserving the required degeneracy, and we shall not write it further on as a part of operating protocols.
(2) Applying an additional voltage $V(t)$, symmetrically over the gate electrodes, we can create approximately homogeneous electric field $\mathcal{E}(t)$ along the central part of the junction where the CPB moves, with a zero-value of the corresponding electrostatic potential in the middle (i.e. approximately preserving the degeneracy of the CPB states). 
While $V_G$ is kept constant, switching the $V(t)$ on and off, as well as $V_b$, is an important part of time-protocols for achieving the desired functionality as will be shown in the following section. $V(t)$ can in principle depend on time to create the time-dependent electric field, yielding different types of nanomechanical motion, but in the present setup to achieve the required functionality we show that it is enough to apply constant additional gate voltage, i.e. $V(t)=V$.

We write the time-dependent Hamiltonian describing the system, with couplings linear in $x$- and $y$-displacement of the CPB, in the form $H(t)=H_0(t)+H_{int}(t)$, where
%
\begin{eqnarray}\label{Hamiltonian}
H_0(t)&=&E_J\sigma_x \cos\Phi(t)+\hbar\omega(a^\dag a+ b^\dag b),\nonumber \\
H_{\text{int}}(t)&=&\epsilon_x \tfrac{1}{\sqrt{2}}(a+a^\dag)\sigma_y \sin\Phi(t)+\epsilon_y \tfrac{1}{\sqrt{2}}(b+b^\dag)\,\sigma_z. \,\,\,\,\,\,\,\,
\end{eqnarray}
%
In Eq.(\ref{Hamiltonian}), $H_0(t)$ represents the noninteracting part of total Hamiltonian $H(t)$, comprising the position-independent part of Josephson tunneling term, characterised by the Josephson energy $E_J$, and 2D harmonic oscillator characterised by frequency $\omega$. Operators $\sigma_{x,y,z}$ are the Pauli matrices operating in the  $2 \times 2$ Hilbert subspace of the CPB, while  $a^\dag$ and $b^\dag$ are phonon creation operators for $x$- and $y$-direction of oscillator motion respectively. $H_{\text{int}}(t)$ is the interacting part of total Hamiltonian, describing position-dependent coupling term of Josephson coupling and coupling of the CPB to gate electrodes, both linear in displacements described by operators $\hat x=a_0 \hat X$ and $\hat y = a_0 \hat Y$,  $\hat X= (a^\dag+a)/\sqrt{2}$ and $\hat Y= (b^\dag+b)/\sqrt{2}$, with corresponding conjugate momenta $\hat p_x=(\hbar/a_0)\hat P_x$ and $\hat p_y=(\hbar/a_0)\hat P_y$, $\hat P_x= i(a^\dag-a)/\sqrt{2}$ and $\hat P_y= i(b^\dag-b)/\sqrt{2}$, respectively. Josephson coupling is characterised by $\epsilon_x \equiv \varepsilon E_J$. Coupling to the gate is $\epsilon_y \equiv 2|e|\mathcal{E}a_0$, where $e$ is an electron charge and $\mathcal{E}$ is electric field created by voltage $V$ on gate electrodes.
In our approach we assume that both couplings are weak, i.e. $\epsilon_x \ll \hbar \omega$ and $\epsilon_y \ll \hbar \omega$.

\section{The time-evolution protocol: implementation of the 3-qubit encoder}

The desired functionality of an encoder for the 3-qubit bit flip code is achieved by operating the external control parameters: bias voltage $V_b(t)$ and gate voltage $\delta V(t)$. Each interval corresponds to the time-protocol described by specific quantum time-evolution operator of the system, governed by the Hamiltonian, which will transfer the system into specific quantum state described by the wave function $\mathbf \Psi(t)$, starting with the given initial state of the system $\mathbf \Psi(t=0)$. We assume that, initially, state of the system is prepared as
%
\begin{eqnarray}\label{Psi_0}
\mathbf \Psi(0)= \mathbf e_{in} \otimes |0\rangle_x \otimes |0\rangle_y,
\end{eqnarray}
%
where $\vert 0 \rangle_x \otimes |0\rangle_y$ is a zero-phonon ground state of mechanical subsystem (the 2D oscillator), while $\mathbf e_{in}$ is the initial charge-qubit state into which the QI is encoded. It is the "QI input" into the system which can be encoded in an arbitrary way, but to deal with it analytically, we choose to work with projections to $\mathbf e_y^\pm$ vectors characterised by coefficients $c_\pm = \mathbf e_y^\pm \mathbf e_{in}$ which will be "bearers" of encoded QI further on. Vectors $\mathbf{e}_i^{\pm}$, $i \in \{ x,y,z \}$ are the eignevectors of $\sigma_i$ Pauli matrices corresponding to eigenvalues $\pm 1$.

The time-evolution operator $\hat U(t,t_0)$ of the wave function from time moment $t_0$ to $t$ is generally solution of the equation $i\hbar \tfrac{\partial}{\partial t}\hat U(t,t_0)=H(t) \hat U(t,t_0)$, with initial condition $\hat U(t_0,t_0)=\mathbf{1}$. It can be found by number of standard methods of quantum mechanics, depending on the properties of $H(t)$, the most popular among them being using the interaction picture representation.\\

Our intention is building the nanomechanical coherent states entangled to charge-qubit state to serve as the ancilla qubits. At $t=0$ we switch on the bias voltage from zero to constant value $V_b$, i.e. $V_b(t)=V_b \Theta(t)$, to provide finite phase difference $\Phi(t)$ in the SC electrodes. As shown in Refs. \cite{npjDR,Bahrova}, coupling of charge-qubit to mechanical subsystem under certain conditions leads to developing of mechanical coherent states. In that respect, in what follows, we restrict our consideration to the resonant case between Josephson and mechanical frequency, i.e. $\Omega=\omega$. The other condition we find to achieve the desired functionality is setting the constant signal $V$ at $t=0$, i.e. $V(t)=V\Theta(t)$, on the gate electrode to provide coupling $\epsilon_y$.
Under these conditions, the Hamiltonian $H(t)$, defined by Eq.(\ref{Hamiltonian}), is periodic in time with period $T=2\pi/\omega$.

We calculate the evolution operator $\hat U(t,t')$ in the interaction picture, i.e.
%
\begin{eqnarray}\label{U1}
\hat U(t,t')=\hat u_0(t)\hat u(t,t')\hat u^\dag_0 (t'),
\end{eqnarray}
%
with boundary condition $\hat u(t,t) = \mathbf 1$, where
%
\begin{eqnarray}\label{u_0}
\hat u_0(t)= \exp\left[-i \omega t (a^\dag a + b^\dag b)- i \alpha\sigma_x \sin \Phi(t) \right],
\end{eqnarray}
%
and $\alpha \equiv E_J/\hbar\omega$ is the scale defined by ratio of Josephson and mechanical energy.
Then, the equation for the evolution operator $\hat u(t,t')$ attains the form
%
\begin{eqnarray}\label{EqOfMotion1}
i \hbar\frac{\partial \hat u(t,t')}{\partial
t}= H_{\text{eff}}(t) \hat u(t,t'),
\end{eqnarray}
%
where $H_{\text{eff}}(t)=H^{(x)}_{\text{eff}}(t)+H^{(y)}_{\text{eff}}(t)$ is defined by
%
\begin{eqnarray}\label{HeffXY}
H^{(x)}_{\text{eff}}(t)&=& \frac{\epsilon_x \sin \Phi(t)}{2}\left(\hat X \cos \omega t + \hat P_x \sin \omega t \right) \nonumber\\
&\times & \left[\left(\sigma_y +i\sigma_z \right)e^{2i \alpha\sin\Phi(t)} + \text{h.c.} \right], \nonumber \\
H^{(y)}_{\text{eff}}(t)&=& \frac{\epsilon_y}{2}\left(\hat Y \cos \omega t + \hat P_y \sin \omega t\right) \nonumber \\
&\times & \left[\left(\sigma_z +i \sigma_y \right)e^{2i \alpha\sin\Phi(t)} +\text{h.c.}\right].
\end{eqnarray}
%
Under the resonance condition, $H_{\text{eff}}(t)$ is a periodic function with the same period $T$ as the original Hamiltonian $H(t)$. As a consequence, the evolution operator $\hat u(t,t')$ has a property
%
\begin{eqnarray}\label{u-periodicity1}
\hat u(t-T,t'-T)=\hat u(t,t'),
\end{eqnarray}
%
leading to
%
\begin{eqnarray}\label{u-periodicity2}
\hat u(NT,0)= \hat u(NT,NT-T) \hat u(NT-T,NT-2T)... \nonumber\\
... \hat u(T,0)=\hat u^N(T,0), \,\,\,\,\,
\end{eqnarray}
%
where $N$ is a positive integer number.

Using the equation of motion (\ref{EqOfMotion1}), we obtain the operator $\hat u(T,0)$ in the form
%
\begin{eqnarray}\label{u(T,0)}
\hat u(T,0)=\mathbf 1 -i \sigma_y \left( \eta_x \hat P_x + \eta_y \hat P_y \right) +{\cal{O}} (\epsilon_{x,y}^2), 
\end{eqnarray}
%
where
%
\begin{eqnarray}\label{etaXY}
\eta_x &=& \pi \frac{\epsilon_x}{\hbar\omega}\left[J_0(2\alpha)-
J_2(2\alpha)\right], \nonumber\\
\eta_y &=& 2\pi \frac{\epsilon_y}{\hbar\omega}J_1(2\alpha),
\end{eqnarray}
%
and $ J_n(2\alpha)$ is the Bessel function of the first kind.
Neglecting the corrections ${\cal{O}} (\epsilon_{x,y}^2)$ higher than linear in $\epsilon_{x}$ and/or $\epsilon_{y}$, using Eqs.(\ref{U1},\ref{u-periodicity2},\ref{u(T,0)}), we obtain the evolution operator of the system for $t=NT$, i.e. for integer number of periods, in the form
%
\begin{eqnarray}\label{U(NT,0)}
\hat U(NT,0) &=& e^{-2i\pi N(a^\dag a+b^\dag b)} \nonumber\\
&\times & \left[\mathbf 1 -i \sigma_y \left( \eta_x \hat P_x + \eta_y \hat P_y \right) \right]^N \nonumber\\
&\simeq & e^{-iN\sigma_y \left(\eta_x \hat P_x + \eta_y \hat P_y\right)} e^{-2i\pi N(a^\dag
a+b^\dag b)}. 
\end{eqnarray}
%
The expression Eq.(\ref{U(NT,0)}) is valid under conditions
$N\epsilon_{i} \sim 1$ and $N\epsilon_{i}^2 \ll 1$, $i \in \{ x,y \}$.

Accordingly to Eq.(\ref{U(NT,0)}), the wave function $\mathbf \Psi (NT) = \hat U(NT,0) \mathbf \Psi (0)$ in the time moment $t=NT$ is obtained in the form
%
\begin{equation}\label{Psi(NT)}
\mathbf \Psi(NT) = c_+\mathbf e_y^+ \otimes \vert Z_x\rangle\otimes \vert Z_y \rangle + c_-\mathbf e_y^- \otimes \vert
-Z_x \rangle \otimes \vert -Z_y \rangle,
\end{equation}
%
where $\vert \pm Z_x \rangle$ and $\vert \pm Z_y \rangle$ are nanomechanical coherent states for $x$- and $y$-direction of oscillator motion, characterised by eigenvalues $Z_x= N \eta_x$ and $Z_y=N \eta_y,$ being proportional to amplitudes, i.e. displacements from the origin of motion.

By virtue of result Eq.(\ref{Psi(NT)}), dividing the arbitrary time interval as $t=NT+\tau$, $\tau < T$, one can find the wave function at the arbitrary moment of time as $\mathbf \Psi(t) = \hat U(NT+\tau, NT) \mathbf \Psi(NT) \simeq \hat u_0(\tau) \mathbf
\Psi(NT)$, yielding the general result
%
\begin{eqnarray}\label{Psi(t)}
\mathbf \Psi(t) &\simeq &  e^{-i \alpha \sigma_x \sin \Phi(\tau)} 
\left[c_+\mathbf e_y^+ \otimes \vert Z_x(\tau)\rangle\otimes \vert Z_y(\tau)\rangle \right. \nonumber\\
&+ & \left.  c_- \mathbf e_y^- \otimes \vert -Z_x(\tau) \rangle \otimes \vert -Z_y(\tau) \rangle \right],
\end{eqnarray}
%
where $Z_{x,y}(\tau)=Z_{x,y}\exp(-i\omega\tau)$ and $\tau = t-NT$, $N=\mathrm{floor}(t/T)$ (an integer part). Note that pair of coherent states $\pm Z_x$, and pair of coherent states $\pm Z_y$ represent 2 states of a qubit each. Therefore, the wave function Eq.(\ref{Psi(t)}) represents the entangled state of three qubits, $\mathbf e_y$, $\vert Z_x \rangle$ and $\vert Z_y \rangle$, i.e. of charge-qubit and two nanomechanical ones, with the QI encoded inside.
Schematically, it is shown in Fig. \ref{Fig-Schematics}(b).
This particular form of the wave function represents the input for the 3-qubit bit flip code, schematically written in the form $c_+ \mid\uparrow \rangle_1 \otimes \vert + \rangle_2 \otimes \vert + \rangle_3 + c_- \mid \downarrow \rangle_1 \otimes \vert - \rangle_2 \otimes \vert - \rangle_3$.

\section{Physical properties of nanomechanical states}

\begin{figure}
\centerline{\includegraphics[width=\columnwidth]{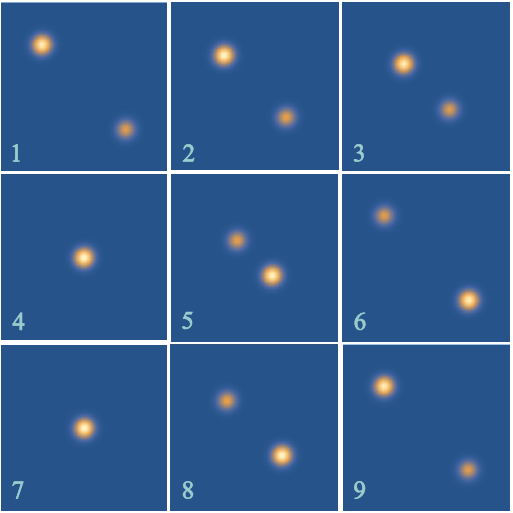}}
\caption{The distribution function Eq.(\ref{Wigner_XY-space}) of the mechanical subsystem in the $(x,y)$-plane of a real space. The scales in pictures are between $-10$ and $10$ in both $x$- and $y$-direction in units $\eta_x$ and $\eta_y$ respectively. Panels 1-9 present the distribution density plot for the time interval $t$ between $5T$ and $6T$ for time moments $\tau=t-NT$, in particular for $\omega\tau  \in \{0,0.75,1,\pi/2,2,\pi,3\pi/2,4,2\pi \}$, $N=5$, respectively, for $|c_+|^2=1/3$, $|c_-|^2=2/3$ and $\epsilon_y/\epsilon_x=1$.}
\label{Fig-Wigner}
\end{figure}

In order to grasp nanomechanical aspects of motion described by the wave function $\mathbf \Psi(t)$, Eq.(\ref{Psi(t)}), i.e. the time-evolution of the mechanical subsystem, it is convenient to calculate the corresponding Wigner function. For that, we need to obtain the reduced mechanical density matrix, $\hat \rho_m(t)=\text{Tr}_q \hat\rho (t)$, where $\hat \rho=\vert \mathbf \Psi(t)\rangle\langle\mathbf \Psi(t)\vert$ is complete density matrix of the system defined by the wave function $\mathbf \Psi(t)$, while by the $\text{Tr}_q$ operation we trace out the charge-qubit degrees of freedom. This operation yields the result 
%
\begin{eqnarray}\label{rho_m}
\hat \rho_m(t) &=& \vert c_+\vert^2\vert Z_x(t)\rangle\langle
Z_x(t)\vert\otimes\vert Z_y(t)\rangle\langle Z_y(t)\vert \nonumber\\
&+& \vert c_-\vert^2\vert -Z_x(t)\rangle\langle -Z_x(t)\vert\otimes\vert -Z_y(t)\rangle\langle -Z_y(t)\vert. \,\,\,\,\, \nonumber\\
\end{eqnarray}
%
The density matrix (\ref{rho_m}) in Wigner's representation (the "Wigner function") is defined as
%
\begin{eqnarray}\label{Wigner_def}
&W&(x,p_x;y,p_y\vert t) = \frac{1}{(2\pi\hbar)^2}\int d\xi_x d\xi_y e^{-\tfrac{i}{\hbar} \left(p_x \xi_x+\imath p_y \xi_y\right)} \nonumber\\
&\times & \langle x+\tfrac{\xi_x}{2}\vert\otimes\langle
y+\tfrac{\xi_y}{2}\vert \,\, \hat\rho_m(t) \,\, \vert x-\tfrac{\xi_x}{2}\rangle\otimes\vert y-\tfrac{\xi_y}{2}\rangle. \,\,\,\,\,\, 
\end{eqnarray}
%
The straightforward calculations (for simplicity we set $\hbar = 1$) yield the result
%
\begin{eqnarray}\label{Wigner}
&&\hspace{-0.5cm}W(x,p_x;y,p_y\vert t)= \nonumber\\
&&\hspace{0.5cm}\frac{\vert c_+\vert^2}{\pi^2}
\exp\left[-\left(x-\mathcal{R}_x(\tau)\right)^2-\left(p_x-\mathcal{P}_x(\tau)\right)^2
\right. \nonumber\\
&&\hspace{2.1cm}- \left. \left(y-\mathcal{R}_y(\tau)\right)^2 -\left(p_y-\mathcal{P}_y(\tau)\right)^2\right] \nonumber\\
&&\hspace{0.2cm}+\frac{\vert c_-\vert^2}{\pi^2}
\exp\left[-\left(x+\mathcal{R}_x(\tau)\right)^2-\left(p_x+\mathcal{P}_x(\tau)\right)^2 \right. \nonumber\\
&&\hspace{2.1cm}-\left. \left(y+\mathcal{R}_y(\tau)\right)^2-\left(p_y+\mathcal{P}_y(\tau)\right)^2\right],
\end{eqnarray}
%
where
%
\begin{eqnarray}\label{Shifts}
\mathcal{R}_x(\tau) &=& \eta_x N\cos\omega\tau, \nonumber\\
\mathcal{P}_x(\tau) &=& -\eta_x N\sin\omega\tau, \nonumber\\
\mathcal{R}_y(\tau) &=& -\eta_y N\cos\omega\tau, \nonumber\\
\mathcal{P}_y(\tau) &=& -\eta_y N\sin\omega\tau
\end{eqnarray}
%
describe the time-dependent shifts from the origin in real ($\mathcal R$) and momentum ($\mathcal P$) space.
Integrating the expression (\ref{Wigner}) over momenta $p_x$ and $p_y$, i.e. $W(x,y\vert t)=\int dp_x dp_y W(x,p_x;y,p_y \vert t)$, one obtains the distribution function that describes evolution of mechanical subsystem in configuration space $(x,y)$, i.e.
%
\begin{eqnarray}\label{Wigner_XY-space}
W(x,y\vert t)&=& \frac{\vert c_+\vert^2}{\pi}
\exp\left[-\left(x-\mathcal{R}_x(\tau)\right)^2-\left(y-\mathcal{R}_y(\tau)\right)^2 \right] \nonumber\\
&+&\frac{\vert c_-\vert^2}{\pi}
\exp\left[-\left(x+\mathcal{R}_x(\tau)\right)^2-\left(y+\mathcal{R}_y(\tau)\right)^2 \right]. \nonumber\\
\end{eqnarray}
%
Analysing the time-dependent shifts Eq.(\ref{Shifts}), it is evident that pair of states $\vert \pm Z \rangle$ represent  linear oscillatory motion in the $(x,y)$-plane who's amplitude grows in time, by $\eta_x$ in $x$- and by $\eta_y$ in $y$-direction, after each period T (see schematic presentation in Fig. \ref{Fig-Schematics}(b) and plots in Fig. \ref{Fig-Wigner}). The inclination of the line of motion in the $(x,y)$-plane is determined by ratio $\epsilon_y/\epsilon_x$. We just mention that by applying the different, time-dependent operating protocols on the gate electrodes, leads to possibility of achieving different types of nanomechanical motion, e.g. circular or Lissajous curves, which is to be discussed elsewhere since it is not crucial for the functionality required here.  
%
%
%
%

\section{Conclusions}

\begin{figure}
\centerline{\includegraphics[width=0.95\columnwidth]{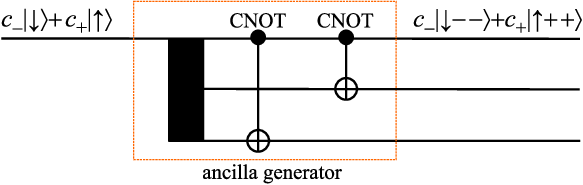}}
\caption{Schematic presentation of the 2-ancilla qubit generator, i.e. the encoding circuit for the bit flip code: $\mid \uparrow \rangle \xrightarrow{\mid + \rangle \mid + \rangle} \vert \uparrow + \, + \rangle$ and $\mid \downarrow \rangle \xrightarrow{\mid - \rangle \mid - \rangle} \vert \downarrow - \, - \rangle$. The one-qubit-state, with encoded QI in the superposition $c_+ \mid \uparrow \rangle + c_- \mid \downarrow \rangle$, gets encoded to $c_+ \vert \uparrow+ \, + \rangle + c_- \vert \downarrow - \, - \rangle$ by action analogous to two (internal) CNOT gates.}
\label{Fig-generator}
\end{figure}

In this paper we propose a concept of functional 2-ancilla qubit generator that, provided the input qubit, builds the 3-qubit input for the bit flip code used in quantum error correction algorithms. In particular, this functionality is implemented in the single physical object, a NEM terminal with an in-plane vibrating superconducting charge-qubit, which is an input qubit, while the two ancilla qubits are nanomechanical coherent states actively created and properly entangled by the very NEM setup. Function of the proposed NEM setup may be presented in terms of mapping $c_+ \mid \uparrow \rangle_1 + c_- \mid \downarrow \rangle_1 \longrightarrow c_+ \mid \uparrow \rangle_1 \vert + \rangle_2 \vert + \rangle_3 + c_- \mid \downarrow \rangle_1 \vert - \rangle_2 \vert - \rangle_3$, where indices represent qubits, 1 - input and 2, 3 - ancillas, while $c_\pm$ are complex coefficients containing the encoded quantum information in qubits' states. Schematically, one may present the function of this setup as an encoding circuit for the bit flip code, operating as two internal CNOT gates, i.e. $\mid \uparrow \rangle \xrightarrow{\mid + \rangle \mid + \rangle} \vert \uparrow + \, + \rangle$ and $\mid \downarrow \rangle \xrightarrow{\mid - \rangle \mid - \rangle} \vert \downarrow - \, - \rangle$ (see Fig. \ref{Fig-generator}).

%
%
%
%
Being easy to manipulate and operate electrically from one end, the negative side of charge qubits, due to the very same  Coulomb interactions with environment, are rather short decoherence times ($10^{-7}$ - $10^{-6}$ s). Among the decoherence processes, the "bit flip", i.e. the charge state fluctuation, appears to be the most frequent and most harmful for the charge-qubit containing the QI. In that respect, "equipping" it with means of increased fidelity or ability of error correction is by all means a useful task.\\

Nanomechanical implementation of the described process, especially on the same physical object, is unique due to its compactness. The NEM setup comprises a cantilever with a mesoscopic superconducting grain on top of it (the charge-qubit), controlled by pair of gate electrodes and pair of biased superconducting electrodes to provide the AC Josephson effect. The coherent states of nanomechanical motion in $x$- and $y$-direction, generated by coupling to electrodes, serve as the "on-board" ancillas for charge-qubit.
\\

The physical feasibility of the proposed setup was mainly discussed in our previous paper on this topic \cite{npjDR} performing number of numerical calculations simulating different decoherence and dephasing processes, mismatch from the resonance condition, control of the applied external voltages etc. Fabrication process of creating the nanopillars (vibrating cantilever - see for example Ref. \cite{CKim}) are heading towards 1GHz operating nanomechanical frequencies with huge quality factors $Q \ge 10^5$. The zero-point amplitude of motion is then of the order of 1-10 pm, tunneling length of the order of 1 - 10 \AA, while the size of the tunneling contacts on terminals is of the order of 10 - 100 nm, within the reach of modern e-beam lithographic techniques.
Bias voltages of the order of $10 \,\, \mu$V is controllable down to $0.1 \% $ (e.g. by Keysight B2961A) which should satisfy the requirements.\\

Investigation of various time-dependent operating protocols on gate electrodes, to induce different types of coupling and corresponding types of controlled nanomechanical motion between contacts in the $(x,y)$-plane, to process quantum information encoded inside, is a subject of our future research.

\section*{Acknowledgements}

This work was supported by the QuantiXLie Centre of Excellence, a project co-financed by the Croatian
Government and European Union through the European Regional Development Fund - the Competitiveness and Cohesion Operational Programme (Grant PK.1.1.02), and by IBS-R024-D1. The authors are grateful to the PCS IBS, Daejeon, Republic of Korea for the hospitality while working on this paper.

\end{document}